\documentclass[manuscript, screen]{acmart}

\usepackage{bm}
\usepackage{subcaption}
\usepackage{tabularray}
\usepackage{xcolor}
\usepackage{ulem}
\usepackage{cleveref}
\usepackage{multirow}
\usepackage{newunicodechar}

\newunicodechar{ệ}{\d{\^{e}}}

\begin{document}

\title{Creator Hearts: Investigating the Impact Positive Signals from YouTube Creators in Shaping Comment Section Behavior}

\author{Frederick Choi}
\author{Charlotte Lambert}
\author{Vinay Koshy}
\author{Sowmya Pratipati}
\author{Tue Do}
\author{Eshwar Chandrasekharan}
\affiliation{
  \institution{University of Illinois at Urbana-Champaign}
  \city{Urbana}
  \state{Illinois}
  \country{USA}
}

\renewcommand{\shortauthors}{Frederick Choi et al.}
\renewcommand{\shorttitle}{Investigating the Impact of Positive Signals from YouTube Creators in Shaping Comment Section Behavior}


\begin{abstract}

Much of the research in online moderation focuses on punitive actions. However, emerging research has shown that positive reinforcement is effective at encouraging desirable behavior on online platforms. We extend this research by studying the ``creator heart'' feature on YouTube, quantifying their primary effects on comments that receive hearts and on videos where hearts have been given. We find that creator hearts increased the visibility of comments, and increased the amount of positive engagement they received from other users. We also find that the presence of a creator hearted comment soon after a video is published can incentivize viewers to comment, increasing the total engagement with the video over time. We discuss the potential for creators to use hearts to shape behavior in their communities by highlighting, rewarding, and incentivizing desirable behaviors from users. We discuss avenues for extending our study to understanding positive signals from moderators on other platforms.
\end{abstract}


\maketitle

\section{Introduction}

\begin{figure}
    \centering
    \includegraphics[width=4in]{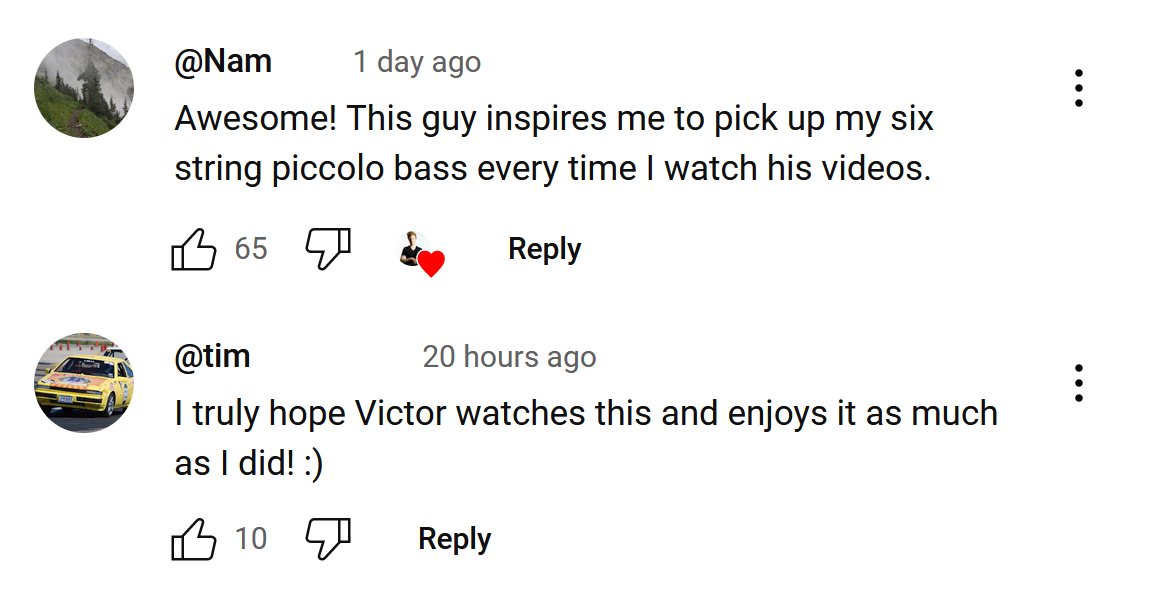}
    \caption{YouTube comments that have received a ``heart'' from a creator are displayed with a distinctive badge (top).}
    \label{fig:heart-example}
\end{figure}

Moderation is an essential part of maintaining a healthy and active online social platform.
However, many of the techniques employed today focus on the moderator's role as a censor or disciplinarian, responsible for eliminating undesirable content and taking punitive actions~\cite{gillespie2018custodians, roberts2016commercial}.
This approach to moderation, while important, leaves several issued unaddressed.
Reactive techniques cannot undo harm already done~\cite{jurgens2019just, xiao2022sensemaking, schoenebeck2023online}, 
and the burden falls on moderators to quickly identify and remove offending content.
On most platforms, effective and timely removal means that the offending content, along with all evidence of the moderator's work, vanish without a trace for most, if not all users.
And so, these moderators who are crucial to the success of online communities and the platforms that host them end up getting little recognition and reward from the community for all the work they do.

The purpose of online moderation and platform governance goes well beyond just censoring content~\cite{grimmelmann2015virtues}.
Desirable activity on a platform is not merely the absence of undesirable activity \cite{bao2021conversations},  and obfuscating or punishing negative behaviors does not necessarily help users understand how to participate in a positive way. 
\citet{jurgens2019just} highlight the approach of articulating affirmative capabilities and promoting positive behaviors instead of trying to articulate and enforce every prohibited behavior.
In addition to removing undesirable content, moderators must establish norms for desirable behavior, and create incentive structures to encourage users to voluntarily adhere to them.

To this end, further research is needed that studies how moderators can use existing affordances and how platform designers can introduce new features to create the incentive structures for desirable behavior.
In this paper, we focus on the moderator's role as a \textit{curator}, responsible not only for stopping undesirable content, but also for promoting desirable content and encouraging desirable behaviors.

\subsection{Creating Incentive Structures to Drive Desirable Behavior}

\citet{kiesler2012regulating} propose several techniques which moderators can leverage to scaffold desirable behavior within their communities. 
Among these, they highlight the importance of positive feedback mechanisms both for rewarding the behavior and for helping others learn the norms by example.

In practice, the effects of positive feedback, particularly positive feedback from moderators, has been difficult to study as clear signals are scarce \cite{lambert2024positive}.
However, by studying formal feedback signals built into the platform (e.g., badges, stars), researchers have been able to investigate their impact on user behavior, the kinds of activities they engage in, and in the content they produce.
\citet{anderson_steering_2013} found that badges on StackOverflow were effective as incentives to steer behavior, reflected in how users would change their behavior to receive badges.
\citet{wang_highlighting_2021} saw a reinforcement effect on comments who had been awarded a New York Times Editors' Pick, exhibiting an increase in both the quality and quantity of their future comments.
It was also shown to have a spill-over effect on other authors who engaged with the comment---the quality of their future comments increased slightly as well.

We extend this line of research to study ``creator hearts'' on YouTube. 

We will use the terms \textit{creators} and \textit{channels} interchangeably in thie paper.
Viewers participate in a community around a creator's content through the comments sections on their videos~\cite{rotman_community_2009}.
Creator hearts allow creators to participate in the community and endorse comments of their choosing by giving them a ``heart''  (illustrated in \Cref{fig:heart-example}).
The unique clarity of creator hearts as signal of endorsement from a creator makes it ideal for studying the effects of positive feedback from distinguished users.
We can examine the causal effects of creator hearts through a quasi-experimental study on YouTube using observational data.
And whereas previous studies of positive feedback were limited in their study populations, studying creator hearts on YouTube allows us to use observational data from a diverse range of creator-audience communities on YouTube.

\subsection{Research Questions}

We investigate how creator hearts can play a role in creating incentive structures for desirable behavior within YouTube comment sections.
To this end, we explore the causal effects of creator hearts---specifically, if and how hearts affect the behavior of the audience community on YouTube---a problem that remains relatively unexplored.
We aim to characterize how creator hearts impact user perceptions and incentives, and investigate if it is worth it for creators to experiment with hearts as a form of positive reinforcement.
We explore how creator hearts affect users' decisions to engage with the comment, as well as the video it was posted on.
Though hearts make comments more likely to be placed at the top of the comments list by design~\cite{Lessard_2016}, we seek to quantify their actual impact on the visibility of comments as well as their ability to draw more attention to the comments that receive them.
Since we are not interested in studying the specific norms of each creator-audience community, we do not analyze the content of the comments themselves.
Instead, we study the activities that users choose to engage or not engage in as a result of receiving or observing a creator heart, which materialize into large scale behaviors that generalize across channels.
Specifically, we ask the following:

\begin{enumerate}
\item[\textbf{RQ1.}] What effect does a creator heart have on how the community engages with the comment that receives it?
        \begin{enumerate}
            \item How does it affect:
            \begin{enumerate}
                \item the placement of the comment in the comment section,
                \item the number of likes the comment receives,
                \item and the number of replies the comment receives?
            \end{enumerate}
            \item What is the relationship of its effect with the size of the community?
    \end{enumerate}
\item[\textbf{RQ2.}] What effect does the presence of creator-hearted comments have on the engagement with the video?
\begin{enumerate}
    \item How does it affect the number of comments the video receives over time?
    \item How do these effects vary based on when the creator heart is given?
\end{enumerate}
\end{enumerate}

\subsection{Summary of Contributions}

\subsubsection{Methods} First, we conducted an initial scan of heart-giving activity across 11.8K channels on YouTube. Then, from a subset of 1K channels, we collected time-series data of comment-level and video-level outcomes for 16.5K videos and 2M comments, and observed the timing of 81.5K creator hearts. We then applied matching and Wilcoxon signed-rank analysis to measure the causal effect of a creator heart on comment-level outcomes (like count, rank, and number of replies) for 11K hearted comments, calculating the results for different channel sizes based on the subscriber count. Finally, we measured the causal effect of the presence and timing of a creator heart on audience participation through comments using a series of regressions on datasets ranging in size from 508-682 videos each.

\subsubsection{Findings}  From our scan of 11.8K channels, we found hearts were given by creators with audiences of all sizes, ranging from 14 to 50M+ subscribers. However, we failed to find any hearts from 44.5\% (5,233) of the channels we scanned. Though we cannot know the exact proportions, it is clear that many, though not all, creators on YouTube are regularly giving away hearts.

We observed that comments that received a heart appeared closer to the start of the comments section. Hearted comments also received an increased number of likes from the community. We found that hearts given within the first 5 hours of the video's publishing had a significant effect on engagement with the video. We observed the greatest effect when a creator heart was given within the first hour, associated with a 22\% increase in the mean number of comments after 12 hours, and a 27.3\% increase after 24 hours.

\subsubsection{Implications} 
We discuss implications for creators in using creator hearts to shape and drive behavior in their communities.
Our findings show that creator hearts can be an effective means of highlighting and drawing the community's attention to comments that the creator wants to endorse. 
Our findings also show that creator hearts are able to incentivise engagement from the audience, and that the earlier a heart is given, the greater its ability to drive engagement.
We discuss how creators can take advantage of these effects to regain some control over how their algorithmically curated comments sections are presented to their audience, and better communicate norms by highlighting examples of desirable behavior. 
We also discuss creator hearts' ability to incentivize participation, the importance of hearts in creating visible traces of creators' engagement within the community, and the implications for similar signals on other platforms in rewarding and incentivizing user behaviors.

Though further research is needed that examines the actual content of the comments, we encourage creators to experiment with using hearts to highlight exemplary comments and to reinforce desirable behavior.

\section{Background}

In this section, we review prior work to situate our research. Specifically, we focus on prior work studying specific interface signals, the shift towards positive forms of feedback, and general research into the YouTube community.

\subsection{Approaches to Online Moderation}

An ever-growing body of research seeks to explain and evaluate the effectiveness of the various moderation strategies employed by online platforms.
At the highest level of governance, platforms have been known to apply sanctions to entire communities by limiting access to them or banning them altogether. 
Studies of such interventions have shown their effectiveness in reducing undesirable behavior from those individuals involved and across the platform as a whole~\cite{chandrasekharan_you_2017, chandrasekharan2022quarantined}.
Many platforms, including Facebook, Instagram, and YouTube, also hire commercial content moderators who manually review and remove user-generated content that can do serious harm to a platform.
However, this work can also do serious harm to the moderator's mental well being~\cite{roberts2016commercial}.

On platforms such as Reddit, Discord, and StackOverflow which are structured around communities, moderators are usually themselves members of the communities they moderate.
These moderators are almost always unpaid volunteers, but are nonetheless vital to the success of online platforms, collectively contributing hundreds of hours worth of moderation work every day carrying out most of the moderation tasks on the platform~\cite{li2022measuring}.
Despite all their work, moderators often struggle to keep up as the scale of their communities and the activity within them grow.
Researchers are continuing to explore the use and development of tools for moderators designed to help them managing their ever increasing workload~\cite{jhaver2019human, chandrasekharan2019crossmod, choi2023convex, kiene2019technological}.

Most of the current practices and research in online moderation, including those mentioned above, focus on moderation through censoring undesirable content. 
While necessary, such measures are insufficient for producing desirable outcomes.
For one, since removing content cannot undo harm already done~\cite{jurgens2019just, xiao2022sensemaking, schoenebeck2023online}, it is important that content is dealt with as quickly as possible---a task that is becoming more difficult as platforms grow.
But, on the other hand, platforms often desire that moderators work invisibly, removing offending content before anyone can see it, and leaving no trace their intervention \cite{roberts2016commercial}.

This lack of visibility of moderator activity is a drawback of removal-focused moderation for several reasons.
First, this lack of transparency can be problematic as it can mislead users to believe the norms they see online reflect norms in the real world.
For example, this can cause a user to experience identity-based harm when 
the content online implies that stereotypes and harmful societal norms related to their identity are the norm offline as well~\cite{simpson2021you}.
Another issue, especially on platforms that rely on volunteer moderators (e.g., Reddit, Discord, StackOverflow), is that the behind-the-scenes efforts of moderators to monitor activity and investigate user-reported incidents go unseen by their communities.
And so, these moderators, who do not receive any compensation from the platform, also end up getting little recognition from their communities for the effort they put in.

Content removals and punitive actions are only one aspect of moderating an online platform~\cite{grimmelmann2015virtues}. 
To achieve a more holistic approach to moderation, we focus this paper on a complementary approach of promoting and encouraging desirable behavior from users on the platform through positive reinforcement and other incentive structures.

\subsection{Signals, Algorithms, and Incentive Mechanisms in Online Social Media Interfaces}

The design of a platform's interface creates incentive structures that influence user behavior.
It is important to study the roles each of the design elements and affordances play in moderating those platforms~\cite{bajpai2022harmonizing}.
Broad design interventions such as reputation systems~\cite{adler_content-driven_2007} and gamification~\cite{hamari_does_2014, sailer_how_2017} have been found to motivate users toward desired actions.
More granular design interventions such as an interstitial in the case of quarantined subreddits have also proved effective in altering the course of user behaviors~\cite{chandrasekharan2022quarantined}.

It is also important to consider how a platform's design affects how users learn norms.
Establishing clear and salient norms is a crucial step for moderating online communities~\cite{kiesler2012regulating}. 
Making injunctive norms more visible, for example, by pinning rules as announcement to report \cite{matias2019preventing}, is shown to increase rule compliance and participation from newcomers.
Making examples of desirable behavior more visible is also important for moderation by influencing how a user learns the descriptive norms.
However, the increasing reliance on algorithmic curation poses a barrier as it becomes harder to control what is shown to users.
This is especially problematic when algorithmic curation promotes the wrong things (or hides the right things), as highlighting too many examples of negative behaviors can lead users to believe that those behaviors are the norm~\cite{kiesler2012regulating}.

The signals a platform implements also play an important role in how users learn norms.
Formal feedback mechanisms are a way to clearly communicate norms by giving users clear signals about how their content is being received by moderators and the community~\cite{kiesler2012regulating}.
Upvotes or likes are a common mechanism for users to give feedback to one another on platforms such as  Reddit, Twitter, and Twitch.
Badges, another example of a positive signal, have been shown to reinforce certain behaviors. 
The New York Times Pick badge was shown to positively correlate with the quality of exposed users' future posts~\cite{wang_highlighting_2021}. 
In the context of Stack Overflow, certain badges can be used to motivate desired behaviors when used strategically~\cite{anderson_steering_2013}. 

It is important to distinguish feedback signals from regular users and moderators since the influence of these signals can depend on the sender's level of authority.
\citet{seering_shaping_2017} explored the imitation effects of Twitch users seeing positive and negative behaviors from different types of users: moderators, subscribers, turbo users, and regular users. 
They found that users with more authority were associated with larger imitation effects for positive behavior.
This prior work establishes a precedent for authoritative figures having the potential to influence change through example-setting. 
However, the research does not account for example-setting or feedback from the Twitch creator.

In our work, we investigate whether the trends found in prior work (e.g., \cite{seering_shaping_2017, wang_highlighting_2021}) 

are present in the YouTube context when considering creator hearts as an explicit signal of approval from an authory figure.
Unlike other work, we do not focus our analysis on moderators, who are the typical figures of authority in online communities. 
While moderators are the users that enforce norms on many platforms, creators uniquely establish the norms and values through their content. Thus, we believe that creators have the potential for strong influence on their communities.

\subsection{YouTube Comments and Communities}

Since its creation, YouTube has allowed viewers to engage with videos by leaving comments visible to the public in a video's comment section.
They also serve to bridge creators and viewers to form creator-audience communities~\cite{rotman_community_2009}.
In 2016, YouTube introduced a feature that allows creators to send positive signals to their community by giving ``hearts'' to comments on their videos~\cite{Lessard_2016}.  
Comments that have received a heart are displayed with a distinctive badge, shown in \Cref{fig:heart-example}. 
They are a formal feedback mechanism for the creator to signal their endorsement.
Contrasting with other formal feedback mechanisms on the platform such as likes and dislikes, creator hearts provide a way for users to receive a signal of approval directly from the single most authoritative figure in their community. 
A heart badge is easy to identify, it is visible to all users, and the source of the heart is unambiguous since only creators can give hearts.
This raises opportunities for us to explore its use and effects in practice.

There have been several studies involving YouTube comments, including explorations into what content gets rated highly~\cite{siersdorfer_how_2010}, what knowledge is shared through commenting~\cite{dubovi_empirical_2020}, and why people comment~\cite{schultes_leave_2013}. 
However, none of these studies touch on how the creator impacts those elements of commenting. 
Additionally, \citet{rotman_community_2009} sought to understand the sense of community on YouTube. Despite the fact that YouTube itself does not have explicit communities, the authors found that YouTube users largely felt like they were part of specific communities. 
The structure of the platform and user interactions do not necessarily define the community structure. Instead, YouTube communities seem to be centered largely around content.
Considering this emphasis on content-based communities, we seek to quantify the role of creator feedback within creator-audience communities on YouTube. 
We expand on prior work exploring quantitative impacts of creator hearts \citet{byun2023effect} by investigating the impact on the comments that receive them, as well as investigating the impact of the timing of when creator hearts are given on videos where they are present.

\section{Data Collection}

To study the effect of creator hearts on comment-level outcomes (RQ1) and video-level outcomes (RQ2),  we construct a dataset of hearted comments from a diverse sample of channels across YouTube. In this section, we describe our data collection process. This involves generating a stream of randomly sampled YouTube channels, scanning channels for hearts previously given, and tracking channels for new hearts. 

We report overview statistics of the data collected in \Cref{tbl:clc-overview}. In \Cref{sec:rq0}, we provide descriptive statistics of the data and briefly characterize overarching patterns of behavior in how creators give out hearts. 

\begin{figure}[]
    \centering
    \includegraphics[width=\textwidth]{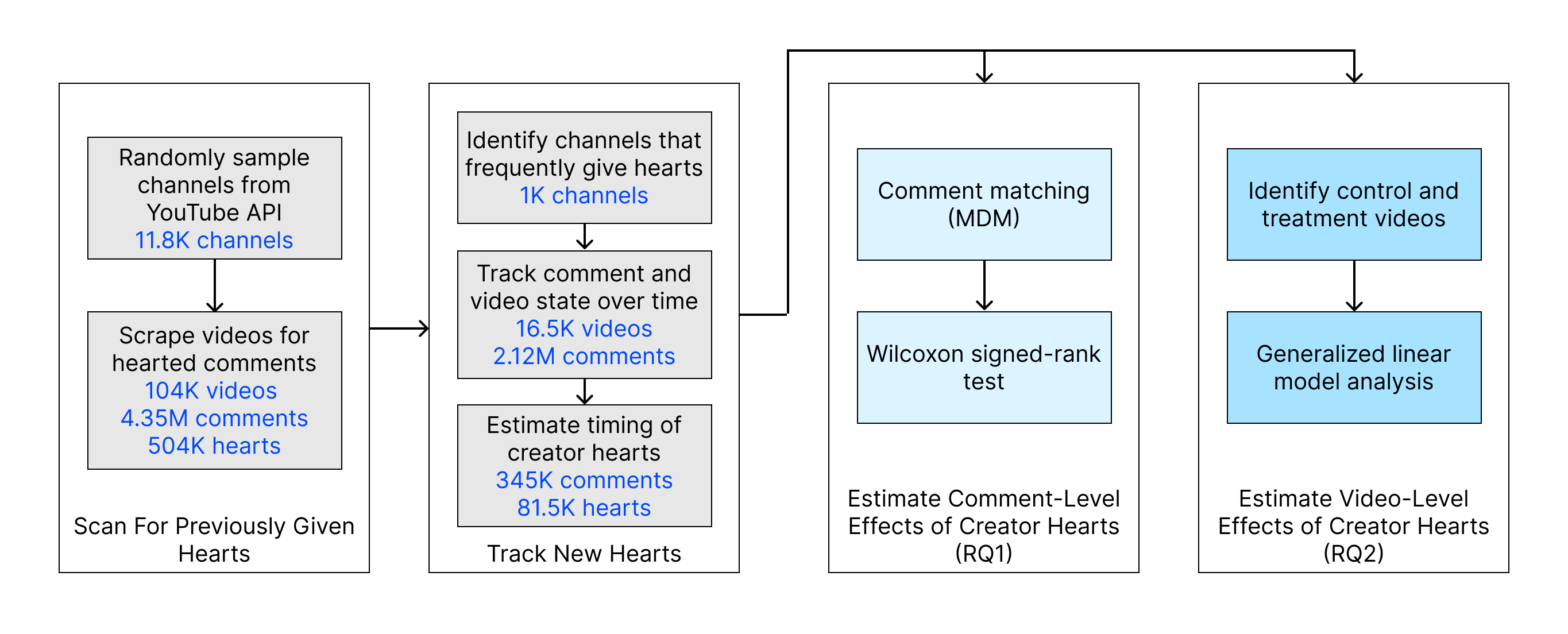}
    \caption{Flowchart depicting the data collection process and subsequent analysis, with statistics of the data collected at each stage.}
    \label{fig:pipeline}
\end{figure}

\subsection{Constructing a Random Sample of Channels}

In the first part of our data collection process, we crawled the Youtube Data API to collect a random sample of videos over time, from which we created a stream of randomly sampled channels to be used for the following stages. Using a script, we queried the YouTube API search endpoint to collect up to 50 of the most recently published videos once every 30 minutes. This initial sampling occurred continuously between February and April 2023.

One limitation of our technique is that our sampling pool is biased towards channels that post more frequently, as their videos would have had more chances to appear in one of our queries. Another limitation stems from the YouTube API itself. Although the API allows us to query for a fixed number of videos published within a specified time frame, it is not clear how the YouTube API selects which videos to return out of the thousands that match. Although the documentation indicates that the ordering of results can be specified to return the most recently published videos first, observation of the returned videos contradicts this. We acknowledge that this introduces unknown biases into our sample. However, since the topics of the videos span dozens of categories and the subscriber counts and average comment counts of the channels in our sample span several orders of magnitude (as shown in \Cref{sec:rq0}), we believe our results will still be generalizable to a broad range of YouTube creators.

\subsection{Scanning for Previously Given Hearts}

In the second part of our data collection process, we scanned for hearted comments on videos from sampled channels. We used this dataset to get a preliminary understanding of the overall presence of heart-giving behavior and trends across creators on YouTube.

Unfortunately, the YouTube API does not provide an easy way to identify creator-hearted comments. We had to instead rely on scraping YouTube's interface. We used Selenium in Python to automate this process, but identifying hearted comments remained slow. Loading the page for a video would take several seconds, even with images disabled, and each group of 10-20 comments would take an additional second or so to load in. To avoid undue strain on the YouTube servers, as well as on our own network bandwidth, we limited our scraper to operate sequentially on only one video at a time.

With this limitation in mind, we narrowed our search for hearted comments to the top 100 comments on the 10 most recent videos for each channel. We performed this procedure starting in March and through April of 2023 while we continued to discover new channels. In total, we collected data for 11.8K channels, and identified 501K hearted comments out of the 4.31M comments we had scraped from 102K videos. 

\subsection{Tracking New Hearts}

To study the effects of a creator heart on comment-level (RQ1) and video-level outcomes (RQ2), we needed to track the comments on each video over time and observe {\it when} comments receive hearts from creators. We used the YouTube API to collect snapshots of comment- and video-level outcomes, and we used our scraper to track which comments had received a heart or not. To increase the likelihood of observing a heart being given, we selected a subset of 1K channels of the 11.8K channels we had previously observed based on how frequently and consistently they had given hearts previously. We then subscribed to the RSS feeds for these channels so that we could start monitoring new videos as quickly as possible. We decided to cycle through up to 20 videos at a time so that consecutive snapshots would be no longer than 5-10 minutes apart on average. After monitoring a video for 4 days, we would retire the video, and its spot in the cycle would be replaced when we received  notification that a new video was published. This process ran between August and December of 2023. A total of 2.12M comments from 16.5K videos were tracked via the API, of which 345K comments were tracked via the scraper. This yielded a total of 82K hearted comments.

\begin{table}
\centering
\caption{Overview statistics of creator hearts data after each stage of data collection. The initial identification of channels was carried out over February-April of 2023, and the tracking was carried out over August-December of 2023.}
\label{tbl:clc-overview}
\begin{tblr}{
  width = \linewidth,
  colspec = {Q[250]Q[100]Q[100]Q[100]Q[100]Q[100]},
  cells = {r},
  row{1} = {c,b},
  cell{2}{1} = {c},
  cell{3}{1} = {c},
  hline{2} = {-}{},
}
\textbf{Stage} & \textbf{Channels} & \textbf{Videos} & \textbf{All Comments} & \textbf{Scraped Comments} & \textbf{Hearts}\\
Totals after scanning for previously given hearts & 11,750 & 102,450 & 4,564,895 & 4,308,795 & 500,620\\
Totals after tracking creator hearts & 1,002 & 16,514 & 2,124,053 & 344,937 & 81,503
\end{tblr}
\end{table}
\subsection{Descriptive Statistics}
\label{sec:rq0}

\begin{figure}
    \centering
    \includegraphics[height=2in]{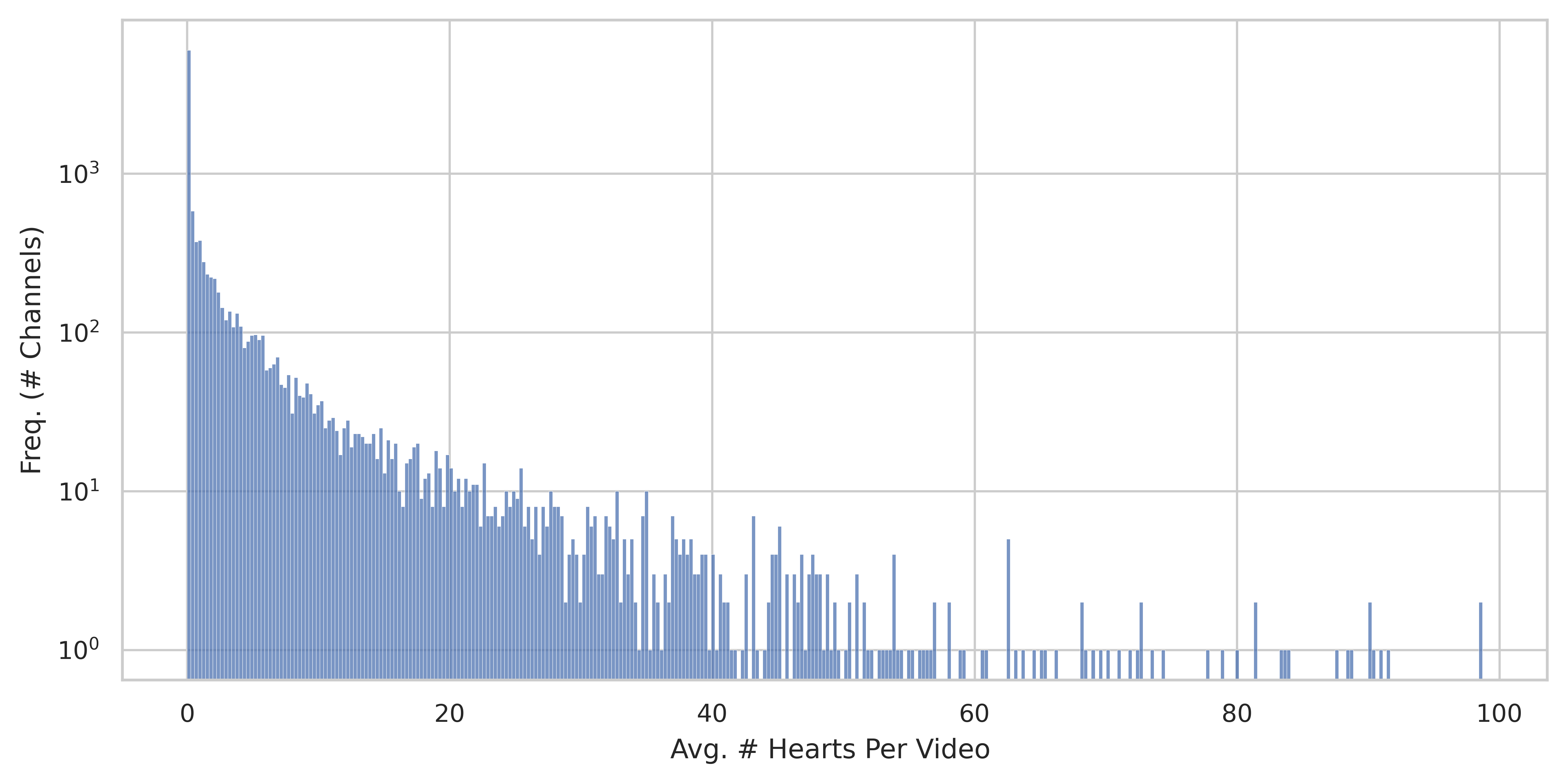}
    \caption{Distribution (across channels) of the average number of creator-hearted comments (of the top 100 comments) per video. We observed at least one heart from 55.5\% (6,517) of the channels in our sample, and an average 6.75 hearts per video from these channels.}
    \label{fig:avg_n_lbc_t100}
\end{figure}

\begin{figure}
    \centering
    \includegraphics[height=2in]{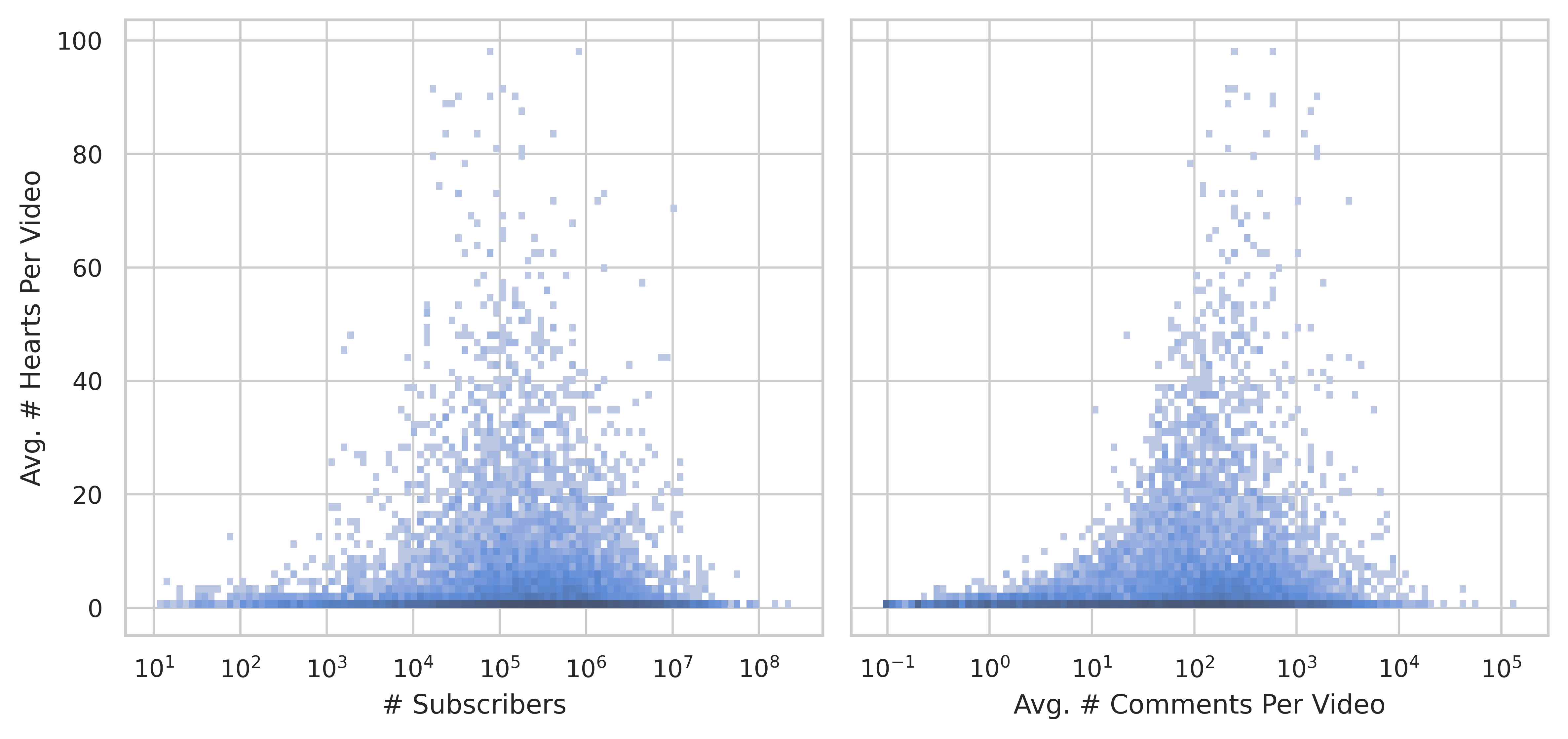}
    \caption{Distributions of the number of hearts observed from each channel, averaged across their videos.}
    \label{fig:2d_avg_n_lbc_t100}
\end{figure}

From the initial scan of 11.8K channels, we found hearts from 55.5\% (6517) of channels. Some of the channels giving out hearts had as few as 14 subscribers, where as others had more than 50M. These numbers demonstrate that giving out hearts is widely practiced by channels of all sizes, though this behavior is hardly universal. 

The distribution of the average number of hearted comments we found across each channels' videos (\Cref{fig:avg_n_lbc_t100}) shows that creators who give out hearts do not just stop at one, but tend to give out several per video. In fact, amongst channels  that give out hearts, we observed an average of 6.75 hearted comments per video.

We found the most likes from creators with between 10K and 1M subscribers and who receive 100-1K comments on average per video (\Cref{fig:2d_avg_n_lbc_t100}). One explanation is that channels (i.e., creators) with few subscribers and who receive few comments on average would be limited in how many hearts they can give. On the other extreme, channels with many subscribers and who receive many comments on average may have given hearts to comments that we did not detect among the comments we scraped. That being said, the typical viewer of those videos are unlikely to see those hearts either, and may get the impression that larger channels are less engaged with their community.

\section{RQ1: Effects on Engagement with Hearted Comment}

After data collection, we address RQ1 and examine the causal effect of a creator heart on comment-level outcomes. This section will first describe the use of Mahalanobis Distance Matching (MDM)~\cite{rubin2006affinely} to match up hearted comments with non-hearted comments that have already received similar amounts of community engagement and have similar potential for future engagement. Then, we detail the results of performing a Wilcoxon signed-rank test to determine what effect creator hearts have on engagement and visibility outcomes.

\subsection{Matching Comments for Causal Inference}

In order to explore whether hearted comments have different outcomes, we first identify a set of control comments and treated comments. The treatment group consists of all comments in our dataset that received a creator heart. Specifically, the treatment group contains the snapshots of treated comments roughly when they were hearted by the creator. 
The control group contains all snapshots of all comments in the dataset that did not receive a heart from the creator at any point. 
Since the control group is a collection of snapshots, most unique comments appear multiple times in the control group at different points in their lifespan. On average, the control group contains roughly 250 snapshots of a comment. 

To perform matching between our set of treated comments and control comments, we measure the Mahalanobis distance between each pair of treated and control comments based on six covariates:
 \begin{enumerate}
     \item \textbf{Comment age:} seconds since the comment was published.
     \item \textbf{Rank:} rank of the comment within the comment section (note: lower rank comments are visually higher up in the interface).

     \item \textbf{Like count:} number of likes on the comment (does not include creator hearts).
     \item \textbf{Video age:} seconds between video publication and comment publication.
     \item \textbf{Number of replies:} the number of replies on the comment.
     \item \textbf{Number of subscribers:} the number of subscribers to the channel that posted the video.
 \end{enumerate}

 \begin{table}[t]
\centering
\caption{This table reports two measures of match quality for our treatment-control matches. First, we report the standardized mean differences (SMDs) between each of our covariates between the treatment and control groups. Then we report the results of a Mann-Whitney U-test on each covariate between the treatment and control groups.}
\label{tab:smd}
\begin{tabular}{lrrc}
\toprule
\multicolumn{1}{c}{\multirow{2}{*}{\textbf{Covariate}}} & \multicolumn{1}{c}{\multirow{2}{*}{\textbf{SMD}}} & \multicolumn{2}{c}{\textbf{Mann-Whitney U-Test}}             \\
\multicolumn{1}{c}{}                           & \multicolumn{1}{c}{}                     & \multicolumn{1}{c}{U} & \multicolumn{1}{c}{$p$-value} \\
\hline
Age                                            & 0.012                                    & 62104846              & 0.238                       \\
Rank                                           & 0.011                                    & 62155772              & 0.282                       \\
Like Count                                & <0.001                                    & 62589306              & 0.817                       \\
Num. Replies                                   & 0.002                                    & 62739207              & 0.867                       \\
Video Age                                      & 0.005                                    & 62372322              & 0.531                       \\
Subscriber Count                               & <0.001                                    & 62563231              & 0.817                      \\
\bottomrule
\end{tabular}
\end{table}

Intuitively, the goal of the matching procedure is to pair treated comments with untreated comments that would have undergone the same trajectory as their treated match, had they been hearted by the creator. To do this, we try to ensure that matched comments:
\begin{enumerate}
    \item Have received similar amounts of community engagement prior to the intervention
    \item Have similar ``potential'' for future engagement
\end{enumerate}

To achieve 1), we match on pre-treatment number of replies, like count, and rank. To achieve 2), we match on video and comment age, as well as the number of subscribers the channel associated with the video has. Age variables are useful proxies for potential engagement since newer comments, and comments posted early in the video's lifespan have likely been seen by a smaller proportion of a video's audience. Similarly subscriber counts are a rough measure of the size of a YouTube channel's community, giving us some insight into the number of people who might potentially come across a comment.

We log-transformed each of the covariates apart from the reply count to encourage the distributions to be more normal. Using these six identified covariates, we use nearest neighbor matching with replacement based on the measured Mahalanobis distance to match each treated comment with one control comment, resulting in 11,196 matches. Out of these matches, there were 5,265 unique control comments. This is partially because we sampled with replacement and otherwise because our control sample contained multiple snapshots of the same comment at different points in time. That means that the same comment may be a good match for one treated comment in one snapshot, and a good match for a different treated comment at a later snapshot.

We note that our matches are not representative of all channels on YouTube, largely because the treated comments dictate the matches. Since the smallest channel in our dataset in which the creator hearted at least one comment has more than 5K subscribers, the matching process discarded any comments posted in significantly smaller channels. Thus, we are unable to make claims about how creator hearts affect engagement in especially small channels.

To measure the quality of these matches, we use two methods. First, we compute the standardized mean difference (SMD) for our six covariates across the two groups. Prior work considers the control and treatment groups to be appropriately balanced if all covariates have SMDs less than 0.25 \cite{saha_causal_2020, kiciman_using_2018}. As shown in Table \ref{tab:smd}, all of our covariates satisfy this condition, thus we consider our matches to be well-balanced. In \Cref{sec:covariate_dists} we provide the full distributions of standardized differences between treatment and control pairs for each covariate.

Second, we perform Mann-Whitney U-tests on each of the covariates used for matching, similar to the approach taken by \citet{chandrasekharan_you_2017}. Table \ref{tab:smd} reports the U statistics and $p$-values for each of these tests, showing that, for each test, we cannot reject the null hypothesis. This implies that there is no significant difference between these six covariates when comparing treatment and control comments.

\begin{table}[t]
    \centering
    \caption{This table reports the T-statistic and p-value for each of the Wilcoxon signed-rank tests we performed (one for each outcome and channel-size pair). Small channels have up to 100K subscribers, medium channels have between 100K and 1M subscribers, and large channels have more than 1M subscribers. The bottom three rows indicate the percentage of matches in which the corresponding outcome for a treated comment is greater than, equal to, or less than its matched control comment.}
    \label{tab:rq2-results}
    \begin{tabular}{l|lll|lll|lll}
\toprule
\textbf{Outcome} & \multicolumn{3}{c}{\textbf{\# Likes}} & \multicolumn{3}{c}{\textbf{Rank}} & \multicolumn{3}{c}{\textbf{\# Replies}} \\
& small & medium & large & small & medium & large & small & medium & large \\
\hline
T & 334780 & 2287101 & 184370 & 1080190 & 8452634 & 697309 & 37254 & 234862 & 7279 \\
p-value & 0.00 & 0.00 & 0.00 & \textbf{0.08} & 0.00 & 0.00 & 0.00 & 0.00 & 0.00 \\
&&&&&&&&&\\
\% Greater & \textbf{40.48 \%} & \textbf{42.28} \% & \textbf{34.58} \% & 41.61 \% & 40.87 \% & 39.34 \% & 6.31 \% & 6.04 \% & 3.63 \% \\
\% Equal & 39.97 \% & 40.51 \% & 48.93 \% & 10.68 \% & 10.68 \% & 10.83 \% & 79.49 \% & 82.36 \% & 89.12 \% \\
\% Less & 19.55 \% & 17.21 \% & 16.49 \% & 47.71 \% & \textbf{48.44} \% & \textbf{49.83} \% & \textbf{14.21} \% & \textbf{11.6} \% & \textbf{7.25} \% \\
\bottomrule
\end{tabular}
\end{table}

\subsection{Results}

With pairs of treated and control comments identified, we then compared the pairs on three different outcomes. To do this comparison, we first identified the age of each comment at the time of its last snapshot in our dataset, which we refer to as its \textit{death}. Within a pair, we look at the snapshots of each comment at the minimum death age between the two comments. This allows us to compare the outcomes of each matched pair after roughly the same amount of time has passed since the age at the time they were matched.

We focus on three outcomes measured at those final snapshots: number of likes, rank, and number of replies. In  \Cref{sec:normality} we check the normality of the differences for each outcome between the treatment and control groups to determine whether a t-test is appropriate. We find that the differences are not normal when considering the raw data, nor after log-transforming the data. Thus, we proceed with a Wilcoxon signed-rank test.

As shown in Table \ref{tab:rq2-results}, we conducted two-sided Wilcoxon signed-rank tests for each outcome and spearated by channel size. Specifically, we consider \textit{small} channels to be those with less than 100K subscribers ($n=2379$), medium channels to have between 100K and 1M subscribers ($n=6804$), and large channels to have more than 1M subscribers ($n=2013$). With these splits, we carried out 9 statistical tests, which generally reveal that there were significant differences between each of the outcomes for treated comments when compared to their paired control comment. This indicates an observed effect of creator hearts on the trajectory of a comment. To determine the direction of this effect, we calculate the percentage of pairs in which the observed outcome for the treated comment was greater than, equal to, or less than the outcome for its control match. These percentages are reported in Table \ref{tab:rq2-results}.

We observe that treated comments tend to end up with more likes than their control counterparts, with the effect being slightly less apparent in large channels.

Evidently, receiving a creator heart encourages other community-members to like the comment. We also see that the final rank of a hearted comment is lower (i.e., in a higher position of the comments section) than that of its matched control comment nearly 50\% of the time.
YouTube itself says that giving a heart to a comment may give the comment a featured position in a preview of the comments section \cite{youtube_faq}, but we are able to show that receiving a creator heart boosts the exposure of a comment in the comment section itself. 
However, this effect is not significant in smaller channels at a threshold of $p < 0.05$, likely due to the smaller number of comments their videos typically receive.
Finally, we see that hearted comments more often end up with fewer replies than those that did not receive the same feedback, an effect that is stronger in small and medium channels.
This is somewhat counter-intuitive to the expectation that creator hearts encourage engagement, however we note that the most common outcome is that both treated and control comments in a pair have equal numbers of replies at their death. 
We can attribute this to the sparse-ness of replies in our dataset.

\section{RQ2: Effects on Engagement with Video}

\newcommand{\Normal}{\text{Normal}}

\newcommand{\initwindow}{w_{\text{init}}}
\newcommand{\follwindow}{w_{\text{foll}}}

\newcommand{\numcominit}{c_{\text{init}}}
\newcommand{\numcomfoll}{c_{\text{foll}}}
\newcommand{\hasclcinit}{I_{\heartsuit}}
\newcommand{\numsubs}{s}

In this section we examine the causal effect of a creator heart on video-level outcomes. First, we describe the outcomes we are measuring, the potential confounds, and the general approach we take in modelling the effect of a creator heart. Then, we describe our data preparation, and how we identify suitable treatment and control videos to compare outcomes. This is followed by a formal definition of the model we used to analyze the effect of a creator heart. We then present our results from measuring the effect creator hearts on video-level outcomes using Bayesian techniques.

\subsection{Method}

For video-level effects of a creator's heart, the outcome we are interested in is the audience's participation through comments, which we measure as the total number of comments on a video. We want to see if the total number of comments a video receives over its lifetime is affected by whether or not the creator gave a heart to any of the comments within the first few hours immediately after the video is published. We begin by observing the comment section for $\initwindow$ hours starting from when the video is published, and check for the presence of a hearted comment by the end of that initial observation window. Since we cannot wait an infinitely long time to count the ``true'' total, we instead measure the number of comments at the end of a follow-up window of $\follwindow$ hours. 

The number of comments in the follow-up window varies greatly from video to video, but is highly correlated with the number comments that were posted in the initial observation window. This is because both counts are influenced by many of the same factors, such as the composition of the channel's typical audience, the performance of the video in YouTube's recommender system, and qualities of the video itself which may encourage more or less participation in the comments. We use the number of comments observed in the initial window as a proxy to control for these confounds that can affect the final comment count. The number of subscribers is another useful covariate to control for varying audience sizes when comparing final counts across different channels. With a regression, we can then see if any of the remaining variation can be explained by the presence of a hearted comment in the initial window.

\subsubsection{Data Preparation}

To make our study more robust, we repeat our analysis with varying initial and follow-up window durations. This has the added benefit of letting us to see how how the creator heart effects change over time. We prepared a total of 12 datasets: 5 with the initial window fixed and varying follow-up windows ($(\initwindow, \follwindow) \in \{1\} \times \{12,15,18,21,24\}$), and an additional 7 cases with the follow-up window fixed and varying initial windows ($(\initwindow, \follwindow) \in \{2, 3, ..., 8\} \times \{12\}$). 

Within each dataset, videos that had at least one hearted comment within the initial window were included as a ``treated'' sample. Meanwhile, videos where all comments posted within the initial window were confirmed (by scraping) to not have been hearted in the initial window were included as a ``control'' sample. Videos where there were unscraped comments posted within the initial window are not included as we cannot reliably determine whether a creator heart was present. We also excluded videos for which we did not have a snapshot of the comment count after the follow-up window since we cannot measure the outcomes. As a result, each dataset consisted of slightly different sets of videos and contained between 508 and 682 videos each.

\subsubsection{Model Definition}

Now, we formalize the model that we used to measure the effect of creator hearts on comment counts within each dataset. Let $\numcominit[i]$ and $\numcomfoll[i]$ be the number of comments on video $i$ at the end of the initial and follow-up windows respectively. Let $\hasclcinit[i]$ be an indicator variable that is equal to one if we observed at least one hearted comment within the initial window of video $i$ and zero otherwise. Let $\numsubs[i]$ be the number of subscribers of the channel that posted video $i$. We utilize a generalized linear model with coefficients $\bm\beta = \{ \beta_0, \beta_\heartsuit, \beta_c, \beta_s \}$ to measure the effect hearted comments ($\beta_\heartsuit$) on the number of comments in the follow-up ($\numcomfoll[i]$). In terms of random variables, we model $\log \numcomfoll[i]$ as being drawn from the following distribution:

\begin{equation}
    \label{eqn:glm}
    \log \numcomfoll[i] \sim \Normal(\beta_0 + \beta_\heartsuit \hasclcinit[i] + \beta_c \log \numcominit[i] + \beta_s \log \numsubs[i], \sigma^2)
\end{equation}

Here, $\sigma^2$ is interpreted as the remaining unexplained variance in $\log \numcomfoll[i]$ after regression over the covariates. We regress on the logarithms of comment counts and subscriber counts to encourage normality and to avoid excess leverage from extreme values. This also allows the model to better capture the non-linear relationship between the initial and follow-up counts. The coefficient $\beta_\heartsuit$ can be interpreted as follows: assuming all other factors are identical, a video with a creator heart in the initial window has the effect of increasing the expected number of comments after the follow-up by a factor of $\exp(\beta_\heartsuit)$. We report effect sizes as the percent increase in the expected number of comments: $100 \times (\exp(\beta_\heartsuit) - 1)$.

\subsubsection{Bayesian Inference}

We use Bayesian techniques to implement the model above~\cite{mcelreath2018statistical}. Instead of estimating a single maximum likelihood value for each parameter, this approach produces distributions for each of the parameters, with likelihoods assigned to a range of possible values for each parameter. In our results, we report several key statistics from the resulting posterior distributions: the mean, standard deviation, and the 95\% credible interval. 

When we report the 95\% credible interval for the effect of a creator heart ($\beta_\heartsuit$), it should be interpreted to mean that there is a 95\% probability that the ``true'' effect of a creator's heart lies in that interval~\cite{hespanhol2019understanding}. The effect of a creator heart should be interpreted as significant (at a significance level of 0.05) when the value of 0 (corresponding to no effect) falls outside the 95\% credible interval~\cite{hespanhol2019understanding}.

The Bayesian approach also requires that we place prior distributions on the parameters, which represent our prior beliefs on the values of each parameter. We selected our priors to minimize our assumptions, placing flat priors on the coefficients $\bm\beta$, and Jeffreys' prior~\cite{jeffreys1946invariant} on the variance $\sigma^2$: $p(\sigma^2)\propto \sigma^{-2}$. 

We ran our regressions with MCMC to produce posterior distributions for each parameter using PyMC~\cite{pymc}. Posterior distributions for $\beta_\heartsuit$ are summarized in \Cref{tab:reg-lbc-multi}, and posterior distributions for all parameters are summarized in \Cref{tab:posterior-all}.

\begin{figure}[H]
    \centering
    \includegraphics[height=2.5in]{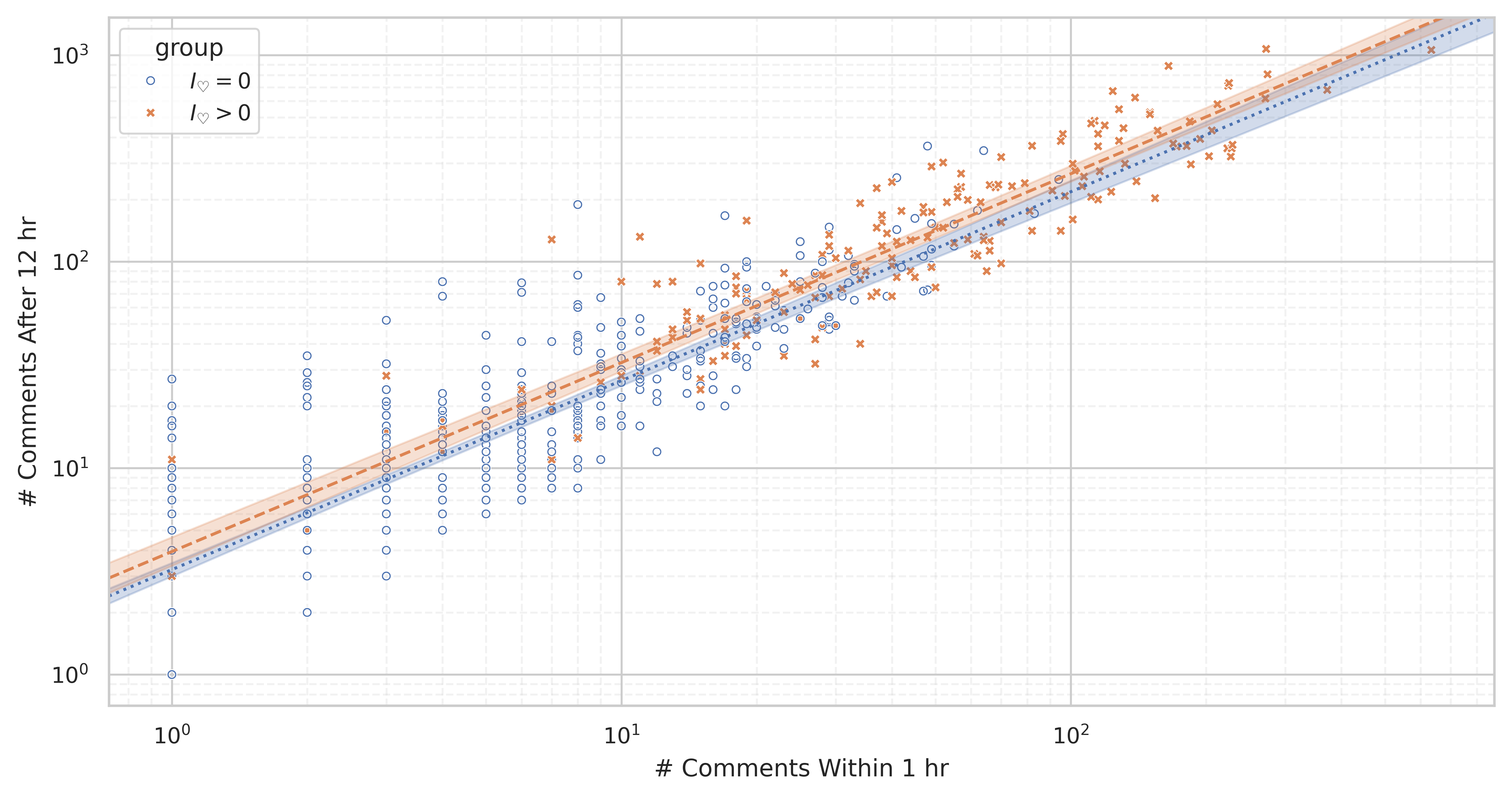}

    \caption{Plot of comment counts from the initial and follow-up windows ($\initwindow=1$, $\follwindow=12$) and posterior means for treatment (dashed line) and control (dotted line) videos. The control group (n=494) consists of videos where no comments were hearted by the creator within the first hour after the video was published ($\hasclcinit = 0$), and the treatment group (n=179) consists of videos where at least one comment was given a heart within the first hour ($\hasclcinit = 1$). The difference in posterior means illustrates a 
    mean 22.1\% increase (HDI 6.3\%-38.7\%) in the number of comments after 13 hours (1 hour initial window + 12 hour follow-up window) from the presence of a creator heart within the first hour.}
    \label{fig:reg-lbc-1h-12h}
\end{figure}

\subsection{Results}

\begin{figure}
    \centering
    \begin{subfigure}{.5\textwidth}
  \centering
    \includegraphics[width=\textwidth]{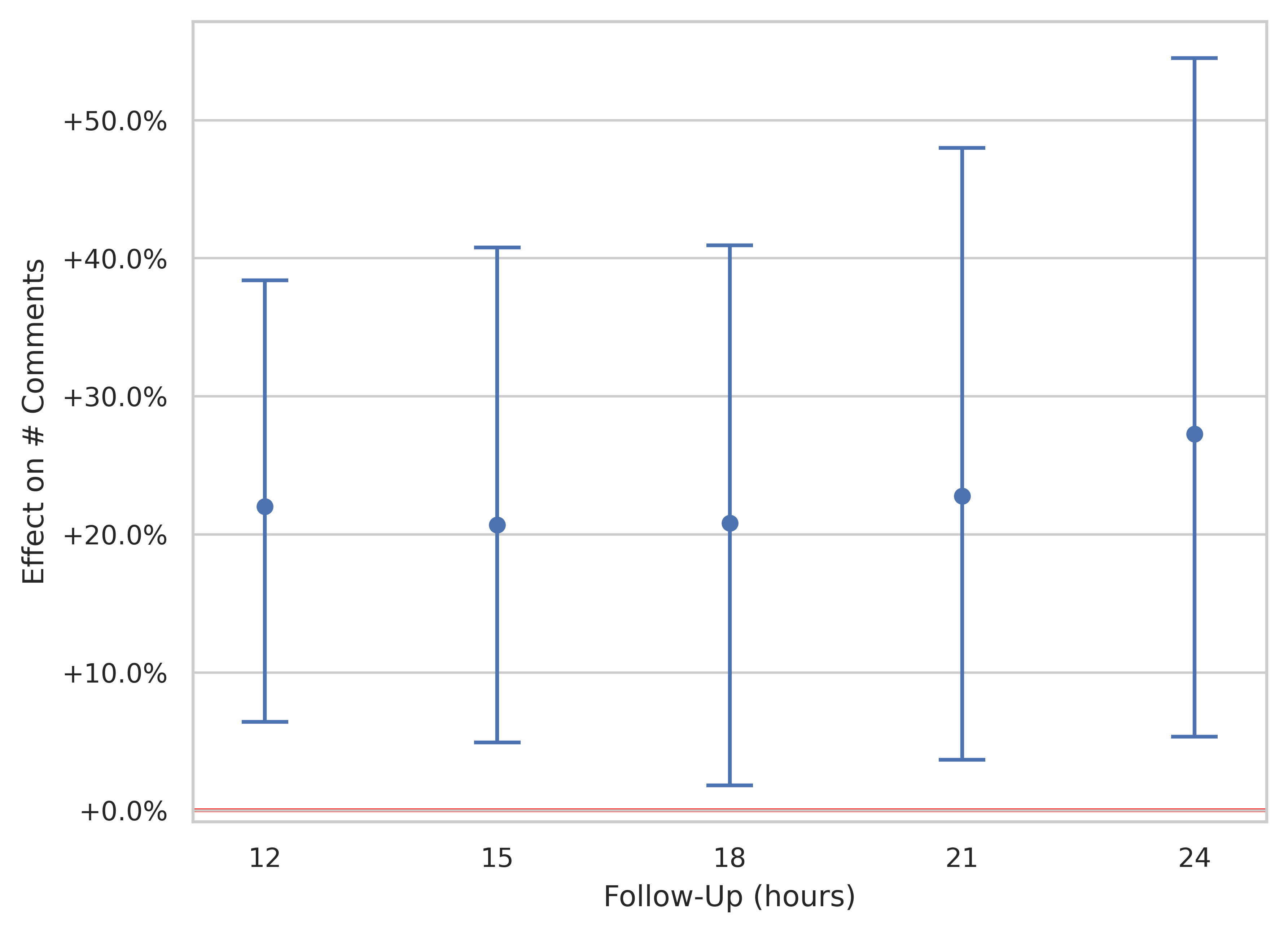}
  \caption{1 hour initial window, varying follow-ups.}
  \label{fig:foll-12-24}
\end{subfigure}%
\begin{subfigure}{.5\textwidth}
  \centering
    \includegraphics[width=\textwidth]{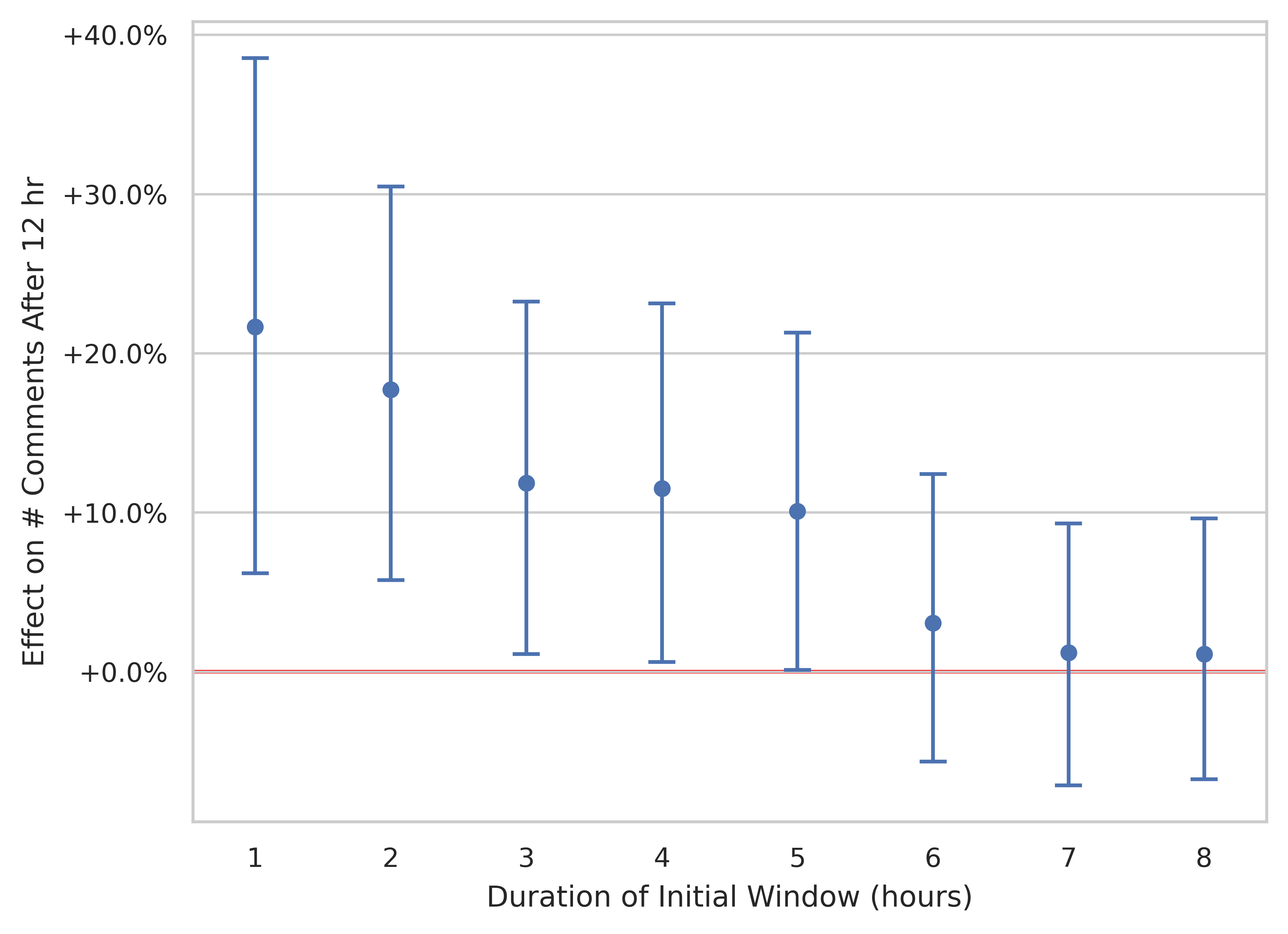}
  \caption{Follow-up after 12 hours, varying initial windows.}
  \label{fig:init-1-8}
\end{subfigure}
    \caption{Effect of the presence of a creator heart within the first few hours of the video publishing. Effects are interpreted as a percent change to the expected number of comments after the follow-up. The markers indicate the posterior means, while the lines indicate the 95\% high-density intervals (see \Cref{tab:reg-lbc-multi} in \Cref{sec:ccpost} for exact values). 
    (a) As the follow-up increases from 12 to 24 hours after the initial window of one hour, the high-density intervals remain greater than zero, indicating that the effect remains significant over time. (b) Meanwhile, as the initial window increases from 1 to 8 hours, the downward trend indicates that the effect of the presence of a creator heart decreases the later it is given.}
    \label{fig:reg-multi}
\end{figure}

Starting with the windows $\initwindow=1$ and $\follwindow=12$, the results show that the presence of a creator heart within the first hour is associated with a mean 22.0\% increase (CI 6.4\%-38.4\%) in the number of comments 12 hours later (\Cref{fig:reg-lbc-1h-12h}). Results from holding the intial window constant while varying the follow-up window from 15-24 hours indicates that the effect remains significant over time, with a mean of up to 27.3\% increase (CI 5.3\%-54.5\%) in the total number of comments after 24 hours (\Cref{fig:foll-12-24}). Together, these results suggest that the presence of a creator heart within the first hour after a video is published has a significant, lasting effect, increasing the audience's overall participation with the video through comments.

To investigate the dependence on when the heart was given, we ran similar regressions with the initial observation period ranging from 1 hour in duration to 8 hours, and following up 12 hours after the end of the initial observation window. The results are illustrated in \Cref{fig:init-1-8}. The downward trend in effect size as the initial observation window increases in duration indicates that the effect of the presence of a creator heart quickly falls off the later it is given, and we were unable to measure a significant effect from an initial window of 6 hours or more. One explanation for the decreased effect size is that the activity in the comments section as a whole might start to level off, diminishing the possible impact a heart can have. Another explanation is that hearts that are given earlier have had more time to be seen by a greater portion of the audience and influence their decision to leave a comment. In either case, the evidence indicates that a creator heart has the greatest impact when it is given within a short window after a video is published.

However, due limitations of the scraper, it was easier to confirm that at least one hearted comment exists (i.e., belongs in the treatment group) than it was confirm that none of the comments had received a heart (i.e., belongs in the control group). 
Since we dropped videos we could not reliably assign a group to, there is a correlation between initial comment count and the group that has been assigned:
The treatment videos in our dataset skew towards having a higher initial comment count and the control videos skew towards a lower initial count. 
It is thus possible that the increase in comments in after the follow-up window actually reflect a more complex relationship between the initial and follow-up counts than our model can predict. 
Future studies can refine the model by accounting for how many views the video had during each window.
This controls for how many users had the opportunity to comment, and arrives at a more direct measure of how creator hearts influenced a user's choice to comment or not.

\section{Discussion}

Based on our findings and prior research, we discuss the ways creator hearts may be useful for shaping user behavior in creator-audience communities.

\subsection{Creator Hearts for Curating Comment Sections}

We found that hearts increase the visibility of the comment that receives it by moving it closer to the start or top of the comment section. 
Our findings also show that a creator heart draws more attention to the comment, evidenced by the increase in the number of likes it receives from the community.
This gives creators the ability to curate their comment sections by directing users' attention to particular comments the creator wants to endorse.

While hearts increase creators' agency over how their comments sections are presented to their audiences, our findings also reveal a tension between the algorithmic curation of comments sections and the content that creators may want to highlight.
Although it is likely that creator-hearted comments are boosted by the YouTube comment sorting algorithm, there are no guarantees that this will always be the case. 
It is possible, for example, that the algorithm might actively deprioritize a creator-hearted comment due to other attributes of the comment (e.g., few user likes). 

This is a fundamental limitation to positive feedback interventions in a social media landscape that is dominated by algorithmically sorted feeds, with implications for platform designers who must weigh curator agency over algorithmic curation.
On YouTube, one way forward might be to afford creators more fine-grained control over how feeds are sorted within their channels. For example, work by \citet{bernstein2023embedding} explores the possibility that higher-level values (e.g. ``pro-democracy'', ``anti-partisanship'') could be embedded into feed-sorting algorithms.
This can be complemented with ways for creators to ``spot-fix'' or override the sorting of individual comments to highlight exemplary behavior, such as by increasing the number of ``pinned'' comments.
Giving creators this level of control could result in an even more powerful norm-setting tool relative to the softer-touch approach of creator hearts, and these tools can be extended to curators on other platforms as well.

\subsection{Incentivizing Participation with Creator Hearts}

We initially hypothesized that hearts would also incentivize users to participate more in the comments section. We found this to be the case---after adjusting for a number of confounders, we estimate that videos where creator likes are deployed will see a 20\% increase in the number of comments observed after a 12-hour follow-up period. Our findings also show that the earlier a creator gives a heart, the greater the impact on engagement with the video. This empirical evidence is valuable to creators, who usually want to increase video engagement in hopes of being favored by the YouTube recommendation algorithm.

This also highlights the value of hearts in allowing creators to leave visible traces of their engagement and activity within a community.
By letting the audience know that rewards are available for those who comment, it is possible that creator hearts incentivize more participation from viewers.
Even for a user who is not looking to receive a creator heart for themselves, they may be more inclined to leave a comment if they believe the creator might read it.

Still, despite the immediate value in the results, the precise mechanism through which positive feedback incentivizes participation is unclear. 
One possibility is that the increased visibility that creator hearts bring may be seen as a reward for users who desire their comment to be seen by others in the community.
Another possibility is that receiving a creator heart is intrinsically rewarding.
Creator hearts may be perceived as granting the receiver a small share of the creator's social capital as they are clear signals of endorsement from them.

Disentangling these mechanisms is essential to understanding the conditions under which positive feedback moderation interventions will be effective. 
For example, if the results are largely explained by the indirect effects of creator hearts and the resulting attention from the community, then it is plausible that similar feedback mechanisms would be effective on many different platforms.
However, if the results are instead largely explained by the creator's own social capital and the relationship between creators and their audience, then we can reasonably assume positive feedback interventions would be similarly effective on a platform like Twitch, but less effective on a platform like Reddit where community moderators are not always known by users.
It is also possible that different mechanisms dominate in different communities within the same platform.
The differential effects of positive feedback through various mechanisms warrant further study, the results of which will be instrumental to the design and integration of more effective incentive structures into the interfaces of online social platforms.

\subsection{Toward Positive Reinforcement} 

Establishing clear and salient norms is essential for moderating online communities~\cite{kiesler2012regulating}. 
The interfaces of many platforms, including YouTube, lack affordances for community leaders and moderators to communicate injunctive norms through explicit rules and guidelines, and providing such affordances would require the introduction of major interface elements.
However, prior work~\cite{wang_highlighting_2021} has shown that highlighting examples of desirable behavior can be an effective way to shape descriptive norms, i.e., what users perceive is the typical or desirable behavior within the community.
Our work demonstrates that the use of creator hearts boosts the visibility of comments, which makes creator hearts effective as a means for highlighting positive examples of participation in the community.
Future work can investigate whether users exposed to these highlighted comments adopt similar behavior in the future. 
Additionally, we were able to show that creator hearts increase participation in a community. Future work can explore whether creator hearts have the power to increase the specific behavior that was awarded the heart from the community as a whole (e.g., prosocial comments, questions, comments discussing specific topics, etc.).

These areas of future work underscore the need to understand the role of positive reinforcement in YouTube communities. 
Positive reinforcement is a concept introduced in the field of psychology by B. F. Skinner~\cite{skinner_operant_1963, ferster_schedules_1957, skinner_science_1965, skinner_recent_1989} in which a stimulus is introduced to reinforce specific behavior. 
Specific methods for employing positive reinforcement has been studied in the context of education~\cite{anderson_engaging_2014, perryer_enhancing_2016}, workplaces~\cite{bradler_employee_2016, armstrong_gamification_2018}, and parenting~\cite{dinkmeyer_parents_1989, steinberg_impact_1992, areepattamannil_parenting_2010}, and have been found effective at encouraging desired behavior.

The idea of positive reinforcement is broadly applicable to many platforms and online communities and has been previously studied in the context of The New York Times comment section~\cite{wang_highlighting_2021}. 
We believe YouTube is a particularly useful fruitful setting for the study of positive reinforcement in online settings. 
In particular, not many platforms have as clear a signal of endorsement as creator hearts, which enable creators to provide positive feedback transparent to users. 
Additionally, while prior work had focused on a single community at a time, YouTube has a diverse range of communities with varied characteristics that enable a broader analysis of the use and effects of positive reinforcement. 
Furthermore, YouTube specifically would benefit from effective methods of positive reinforcement because of the extent to which comments on YouTube affect audiences and creators themselves. 
\citet{schultes_leave_nodate} found that the comments on a video influence viewers' perceptions of the video, and the number of likes/dislikes it receives.
In addition, \citet{thomas_2022_its} found that 95\% of their surveyed YouTube creators had experienced at least one incident of harassment, many in the form of hateful comments.
These motivate the need for ways creators can effectively reinforce behavior that they want to encourage and shape their communities into more valuable and welcoming environments for both creators and audiences.
With this work, we lay the groundwork for such studies on the impact of creator hearts and positive reinforcement on the development of community-specific behaviors and norms.

\section{Conclusion}

Moderation involves more than just punitive actions; as we move towards supporting moderators in their broader role as curators, there is a need to study more ways to positively reinforce desirable behavior.
In this paper, we quantified the causal effects of \textit{creator hearts} on comments that receive hearts, and on videos where hearts have been given.
We found that creator hearts increase the visibility of comments and increase the number of likes a comment receives from the community.
We found an increase in the number of comments on videos where a creator heart was present shortly after it is published.
We conclude that giving hearts is an easy option for creators who want to highlight exemplary behavior and incentivize participation. 
Our work as well as future research on similar signals in other platforms will be instrumental to the design of interfaces that better support moderators and their active roles in shaping online communities. 

\bibliographystyle{ACM-Reference-Format}
\bibliography{bibliography}

\newpage

\appendix
\section{Full Distributions of Matching Covariates}\label{sec:covariate_dists}

\begin{figure}[H]
  \includegraphics[width=\textwidth]{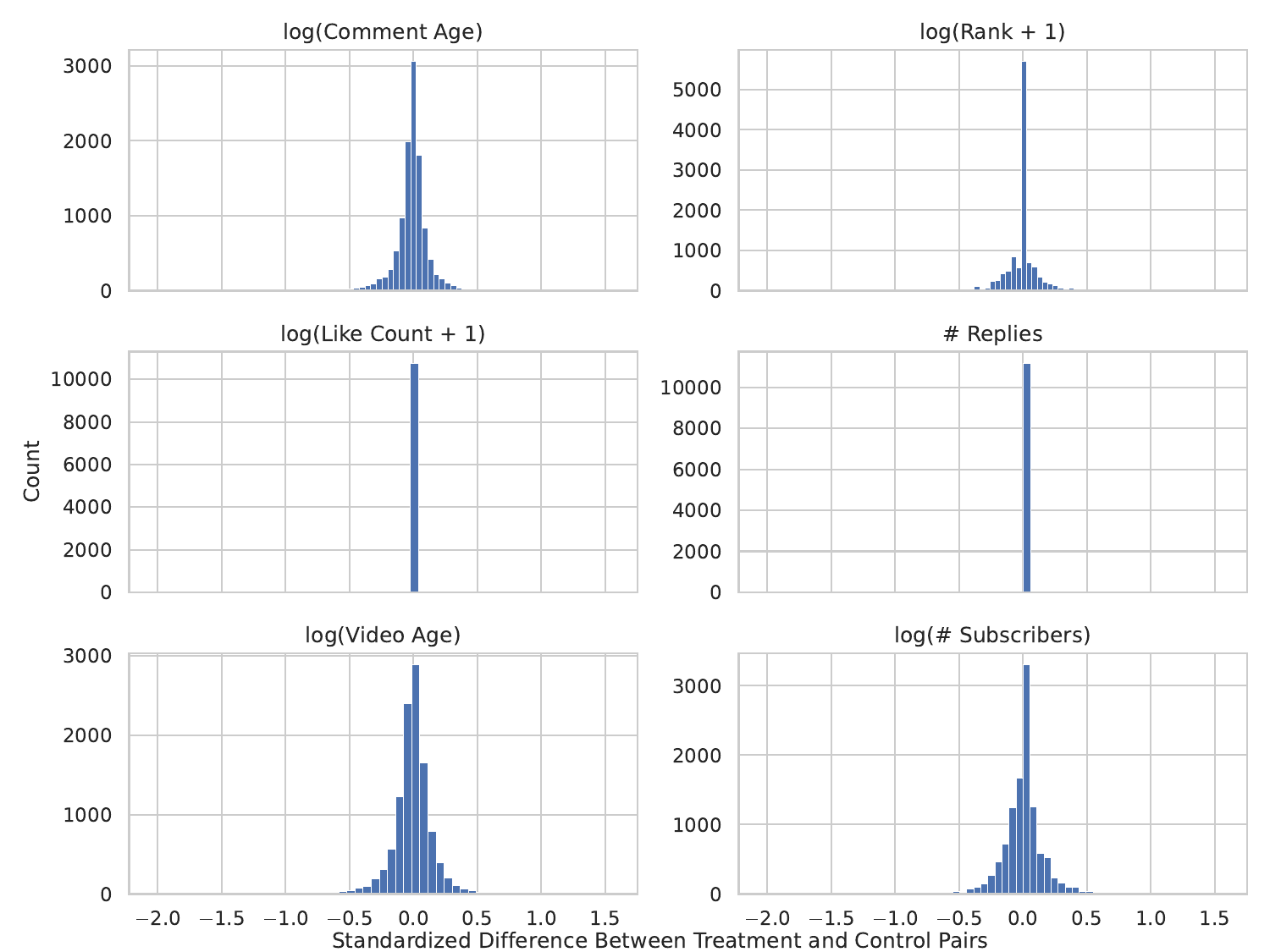}
  \caption{Standardized differences between matched pairs of treatment and control units for each covariate. All plots are roughly centered around zero and relatively concentrated, indicating good match quality }
  \label{fig:triptych}
\end{figure}

\section{Probability Plots to Check Normality}
\label{sec:normality}

\begin{figure}[H]
    \centering
    \includegraphics[width=0.95\linewidth]{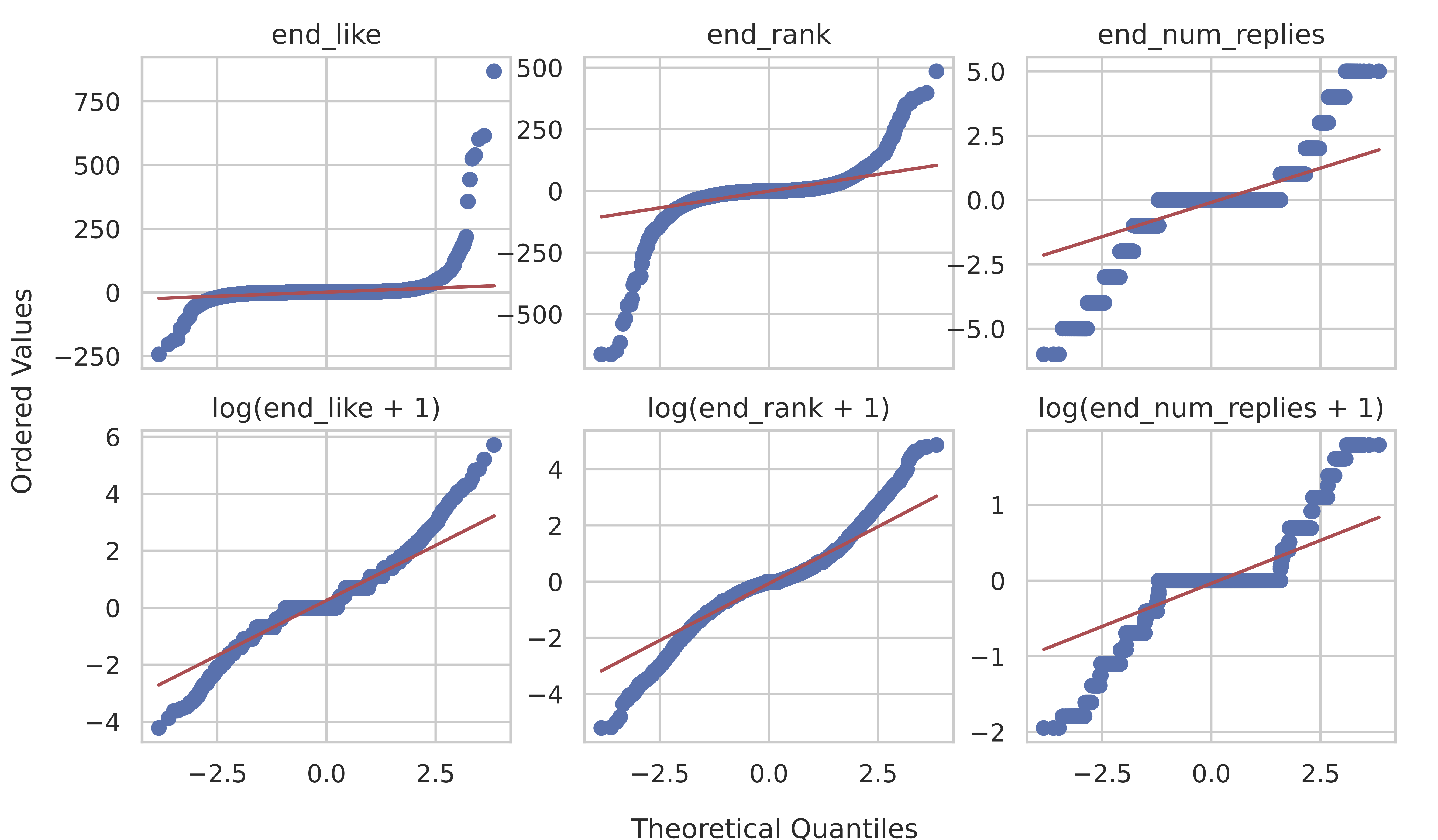}
    \caption{Probability plots of the differences between treatment and control comments for each of our three outcomes against the quantiles of a normal distribution. The red line shows the best-fit line for the data, demonstrating the non-normality of the data.}
    \label{fig:probplots}
\end{figure}

\section{Posterior Distributions of Regressions on Comment Counts}
\label{sec:ccpost}

\begin{table}[H]
\caption{Posterior distributions of $\beta_\heartsuit$ for varying initial observation windows and follow-ups. In parenthesis are the effect sizes interpretated as a percent change to the expected number of comments after the follow-up period, computed as $100 \times (\exp \beta_\heartsuit - 1)$. Significant results are in bold. Results are considered significant when 0 (no effect) falls outside the 95\% credible interval. Results where 0 falls outside the 99\% credible interval are indicated with an additional *.}
\label{tab:reg-lbc-multi}
\begin{tblr}{
  width = \linewidth,
  colspec = {Q[150]Q[300]Q[200]Q[100]Q[400]},
  cells = {c},
  row{1} = {b},
  hline{1,14} = {-}{0.08em},
  hline{2} = {-}{0.05em},
}
{\textbf{Window Durations}\\($\initwindow$, $\follwindow$)} & {\textbf{Sample Size}\\(Total, $\hasclcinit=0$, $\hasclcinit=1$)} & \textbf{Mean} & \textbf{Std. Dev.} & \textbf{95\% CI}\\
1, 12 & 672, 493, 179 & \textbf{0.199 (+22.0\%)*} & 0.067 & {[}0.062, 0.325] ([+6.4\%, +38.4\%])\\
1, 15 & 613, 470, 143 & \textbf{0.188 (+20.7\%)} & 0.075 & {[}0.048, 0.342] ([+4.9\%, +40.8\%])\\
1, 18 & 562, 442, 120 & \textbf{0.189 (+20.8\%)} & 0.084 & {[}0.018, 0.343] ([+1.8\%, +40.9\%])\\
1, 21 & 535, 427, 108 & \textbf{0.205 (+22.8\%)} & 0.091 & {[}0.036, 0.392] ([+3.7\%, +48.0\%])\\
1, 24 & 508, 412, 96 & \textbf{0.241 (+27.3\%)} & 0.099 & {[}0.052, 0.435] ([+5.3\%, +54.5\%])\\
2, 12 & 682, 483, 199 & \textbf{0.163 (+17.7\%)*} & 0.054 & {[}0.056, 0.266] ([+5.8\%, +30.5\%])\\
3, 12 & 672, 478, 194 & \textbf{0.112 (+11.9\%)} & 0.051 & {[}0.011, 0.209] ([+1.1\%, +23.2\%])\\
4, 12 & 668, 475, 193 & \textbf{0.109 (+11.5\%)} & 0.051 & {[}0.006, 0.208] ([+0.6\%, +23.1\%])\\
5, 12 & 656, 469, 187 & \textbf{0.096 (+10.1\%)} & 0.049 & {[}0.001, 0.193] ([+0.1\%, +21.3\%])\\
6, 12 & 641, 464, 177 & 0.030 (+3.0\%) & 0.045 & {[}-0.058, 0.117] ([-5.6\%, +12.4\%])\\
7, 12 & 631, 461, 170 & 0.012 (+1.2\%) & 0.041 & {[}-0.074, 0.089] ([-7.1\%, +9.3\%])\\
8, 12 & 623, 459, 164 & 0.011 (+1.1\%) & 0.041 & {[}-0.070, 0.092] ([-6.8\%, +9.6\%])
\end{tblr}
\end{table}

\clearpage

\begin{longtblr}[
  label = {tab:posterior-all},
  caption = {Posterior distributions of the parameters of the model defined in \cref{eqn:glm} for varying durations of initial and follow-up windows (measured in hours).},
  entry = none,
]{
  width = \linewidth,
  stretch = 0,
  colspec = {Q[150]Q[300]Q[100]Q[100]Q[100]Q[200]},
  cells = {c},
  row{1} = {b},
  cell{2}{1} = {t},
  cell{2}{2} = {t},
  cell{7}{1} = {t},
  cell{7}{2} = {t},
  cell{12}{1} = {t},
  cell{12}{2} = {t},
  cell{17}{1} = {t},
  cell{17}{2} = {t},
  cell{22}{1} = {t},
  cell{22}{2} = {t},
  cell{27}{1} = {t},
  cell{27}{2} = {t},
  cell{32}{1} = {t},
  cell{32}{2} = {t},
  cell{37}{1} = {t},
  cell{37}{2} = {t},
  cell{42}{1} = {t},
  cell{42}{2} = {t},
  cell{47}{1} = {t},
  cell{47}{2} = {t},
  cell{52}{1} = {t},
  cell{52}{2} = {t},
  cell{57}{1} = {t},
  cell{57}{2} = {t},
  hline{1,62} = {-}{0.08em},
  hline{2} = {-}{0.05em},
  hline{7,12,17,22,27,32,37,42,47,52,57} = {-}{},
}
{\textbf{Window Durations}\\($\initwindow$, $\follwindow$)} & {\textbf{Sample Size}\\(Total, $\hasclcinit=0$, $\hasclcinit=1$)} & \textbf{Parameter} & \textbf{Mean} & \textbf{Std. Dev.} & \textbf{95\% CI}\\
1, 12 & 672, 493, 179 & $\beta_0$ & 1.749 & 0.187 & {[}1.375, 2.113]\\
 &  & $\beta_\heartsuit$ & 0.199 & 0.067 & {[}0.062, 0.325]\\
 &  & $\beta_c$ & 0.916 & 0.020 & {[}0.874, 0.954]\\
 &  & $\beta_s$ & -0.045 & 0.015 & {[}-0.075, -0.017]\\
 &  & $\sigma$ & 0.576 & 0.015 & {[}0.545, 0.606]\\
1, 15 & 613, 470, 143 & $\beta_0$ & 1.755 & 0.200 & {[}1.369, 2.148]\\
 &  & $\beta_\heartsuit$ & 0.188 & 0.075 & {[}0.048, 0.342]\\
 &  & $\beta_c$ & 0.911 & 0.022 & {[}0.872, 0.954]\\
 &  & $\beta_s$ & -0.042 & 0.016 & {[}-0.072, -0.012]\\
 &  & $\sigma$ & 0.598 & 0.017 & {[}0.566, 0.635]\\
1, 18 & 562, 442, 120 & $\beta_0$ & 1.858 & 0.216 & {[}1.435, 2.266]\\
 &  & $\beta_\heartsuit$ & 0.189 & 0.084 & {[}0.018, 0.343]\\
 &  & $\beta_c$ & 0.908 & 0.023 & {[}0.861, 0.954]\\
 &  & $\beta_s$ & -0.046 & 0.017 & {[}-0.080, -0.015]\\
 &  & $\sigma$ & 0.613 & 0.018 & {[}0.577, 0.648]\\
1, 21 & 535, 427, 108 & $\beta_0$ & 1.977 & 0.225 & {[}1.547, 2.425]\\
 &  & $\beta_\heartsuit$ & 0.205 & 0.091 & {[}0.036, 0.392]\\
 &  & $\beta_c$ & 0.904 & 0.025 & {[}0.855, 0.953]\\
 &  & $\beta_s$ & -0.053 & 0.018 & {[}-0.088, -0.019]\\
 &  & $\sigma$ & 0.628 & 0.019 & {[}0.591, 0.666]\\
1, 24 & 508, 412, 96 & $\beta_0$ & 2.073 & 0.248 & {[}1.564, 2.517]\\
 &  & $\beta_\heartsuit$ & 0.241 & 0.099 & {[}0.052, 0.435]\\
 &  & $\beta_c$ & 0.901 & 0.026 & {[}0.850, 0.952]\\
 &  & $\beta_s$ & -0.057 & 0.019 & {[}-0.093, -0.019]\\
 &  & $\sigma$ & 0.651 & 0.021 & {[}0.611, 0.693]\\
2, 12 & 682, 483, 199 & $\beta_0$ & 1.072 & 0.148 & {[}0.772, 1.356]\\
 &  & $\beta_\heartsuit$ & 0.163 & 0.054 & {[}0.056, 0.266]\\
 &  & $\beta_c$ & 0.924 & 0.015 & {[}0.895, 0.954]\\
 &  & $\beta_s$ & -0.017 & 0.011 & {[}-0.040, 0.004]\\
 &  & $\sigma$ & 0.480 & 0.013 & {[}0.456, 0.507]\\
3, 12 & 672, 478, 194 & $\beta_0$ & 0.697 & 0.136 & {[}0.434, 0.963]\\
 &  & $\beta_\heartsuit$ & 0.112 & 0.051 & {[}0.011, 0.209]\\
 &  & $\beta_c$ & 0.942 & 0.014 & {[}0.916, 0.969]\\
 &  & $\beta_s$ & -0.003 & 0.010 & {[}-0.023, 0.018]\\
 &  & $\sigma$ & 0.426 & 0.012 & {[}0.404, 0.451]\\
4, 12 & 668, 475, 193 & $\beta_0$ & 0.688 & 0.137 & {[}0.421, 0.954]\\
 &  & $\beta_\heartsuit$ & 0.109 & 0.051 & {[}0.006, 0.208]\\
 &  & $\beta_c$ & 0.936 & 0.013 & {[}0.910, 0.962]\\
 &  & $\beta_s$ & -0.008 & 0.011 & {[}-0.028, 0.013]\\
 &  & $\sigma$ & 0.425 & 0.012 & {[}0.402, 0.449]\\
5, 12 & 656, 469, 187 & $\beta_0$ & 0.611 & 0.133 & {[}0.368, 0.889]\\
 &  & $\beta_\heartsuit$ & 0.096 & 0.049 & {[}0.001, 0.193]\\
 &  & $\beta_c$ & 0.933 & 0.013 & {[}0.907, 0.957]\\
 &  & $\beta_s$ & -0.006 & 0.010 & {[}-0.027, 0.013]\\
 &  & $\sigma$ & 0.405 & 0.011 & {[}0.382, 0.426]\\
6, 12 & 641, 464, 177 & $\beta_0$ & 0.519 & 0.121 & {[}0.274, 0.741]\\
 &  & $\beta_\heartsuit$ & 0.030 & 0.045 & {[}-0.058, 0.117]\\
 &  & $\beta_c$ & 0.960 & 0.012 & {[}0.937, 0.983]\\
 &  & $\beta_s$ & -0.007 & 0.009 & {[}-0.025, 0.011]\\
 &  & $\sigma$ & 0.372 & 0.010 & {[}0.352, 0.392]\\
7, 12 & 631, 461, 170 & $\beta_0$ & 0.460 & 0.111 & {[}0.234, 0.672]\\
 &  & $\beta_\heartsuit$ & 0.012 & 0.041 & {[}-0.074, 0.089]\\
 &  & $\beta_c$ & 0.969 & 0.011 & {[}0.949, 0.991]\\
 &  & $\beta_s$ & -0.007 & 0.009 & {[}-0.023, 0.010]\\
 &  & $\sigma$ & 0.337 & 0.010 & {[}0.317, 0.354]\\
8, 12 & 623, 459, 164 & $\beta_0$ & 0.487 & 0.113 & {[}0.251, 0.698]\\
 &  & $\beta_\heartsuit$ & 0.011 & 0.041 & {[}-0.070, 0.092]\\
 &  & $\beta_c$ & 0.966 & 0.011 & {[}0.946, 0.988]\\
 &  & $\beta_s$ & -0.010 & 0.009 & {[}-0.026, 0.009]\\
 &  & $\sigma$ & 0.333 & 0.010 & {[}0.314, 0.352]
\end{longtblr}

\end{document}